# Adding a Storage Pool improves 3-PG Tree-ring Simulations

Yanfang Wang[1], Liang Wei[1*], Liheng Zhong[2], Xizi Yu[1], Fang Wang[3], John D. Marshall[4,5]

1. MOE Key Laboratory of Western China's Environmental System, College of Earth and Environmental Sciences, Lanzhou University, Lanzhou, Gansu 730000, China

2. Ant Group, World Financial Center, Beijing 100020, China

3. School of Geographical Sciences, Hebei Normal University, Shijiazhuang 050024, China

4. Department of Geosciences, University of Gothenburg, Gothenburg 40530, Sweden

5. Department of Matters and Energy Fluxes, Global Change Research Institute CAS, Brno 603 00 Czechia

* Correspondence: liangwei@alumni.uidaho.edu

**Abstract**

Tree rings provide long-term records of tree growth and climate changes, which makes them ideal benchmarks for forest modeling. Tree-ring information has greatly improved the reliability of 3-PG, which is one of the most commonly used process-based forest growth models. Here we strengthen 3-PG's ability to simulate tree-ring width and carbon stable isotopes ($\delta^{13}C$) by enhancing its descriptions of tree physiology. The major upgrade was adding a carbon storage pool for tree-ring formation using stored carbohydrates. We also incorporated previous modifications (replacing the age modifier with a height modifier) of 3-PG and tested their efficacy in improving tree-ring simulations. We ran the model based on two grand fir (*Abies grandis*) stands. The updated model greatly improved the simulations for both tree-ring widths and $\delta^{13}C$. The results represent one of the best tree-ring $\delta^{13}C$ simulations, which accurately captured the amplitude in annual variations of $\delta^{13}C$. The correlations ($R^2$) between simulations and observations reached 0.50 and 0.73 at two stands respectively. The new model also greatly improved the simulations of raw tree-ring widths and detrended ring-widths index. Because of better descriptions of tree physiology and more accurate simulations of tree rings than the previous model version, the updated 3-PG should provide more reliable simulations than previous 3-PG versions when tree-ring information is used as benchmark in future studies.

# 1. Introduction

Process-based forest models rely on observed forest growth and physiological data for parameterization and validation (Aber, 1997). Most modeling practices have relied on short-term (i.e. minutes, hours, or several years) physiological observations as long-term physiological data are scarce (e.g. Rasse et al., 2001; Weng and Luo, 2011). This may limit our ability to simulate long-term forest dynamics especially when the climate has changed over the past several decades and tree growth may have changed accordingly. It would hence be ideal to have long-term observations (decades or longer) to calibrate and validate models (Jeong et al., 2021). Tree rings provided a convenient data source for model benchmarking as they record long-term physiological and climatic information and can be precisely dated (Evans et al., 2022; Guiot et al., 2014; Jeong et al., 2021; Walcroft et al., 1997; Wei et al., 2022).

Both tree-ring widths and stable isotopes can be used to validate, constrain, and benchmark these models (Barichivich et al., 2021; Jeong et al., 2021; Wei et al., 2022). First, simulating tree-ring stable carbon ($\delta^{13}C$) and oxygen ($\delta^{18}O$) isotopes require accurate descriptions of photosynthetic gas-exchange. This is because gas exchange processes are fundamental to determine plants' $\delta^{13}C$ (Farquhar model; Farquhar et al., 1982) and $\delta^{18}O$ (Craig-Gordon model; Craig and Gordon, 1965). The $\delta^{13}C$ in plants is mainly determined by stomatal conductance and the rate of photosynthesis, both of which are influenced by environmental conditions such as light, temperature, humidity, and $CO_2$ concentration. Besides source water, the $\delta^{18}O$ in the plant is mainly influenced processes of transpiration, which are also determined by air humidity and other environmental factors that changes the transpiration (light, temperature, etc.). Second, reasonably simulating tree-ring widths requires accurate descriptions of not only gas exchange but also carbon allocation. Once carbon is fixed through photosynthesis, it is then allocated to different parts of the tree, including roots, leaves, stems, and reproductive structures. Only when allocation to the stem (where tree rings form) is accurately simulated can tree-ring width be estimated correctly. Third, as environmental conditions change both gas exchange and carbon allocation (Franklin et al., 2012; Wang et al., 2021), models need to account for impacts of variations of environmental factors on the tree phenology and growth to simulate tree-ring stable isotopes and widths. Therefore, the ability to reasonably simulate $\delta^{13}C$ and ring widths can be testament to a model's sophistication in capturing the complex interplay between photosynthesis, stomatal conductance, carbon allocation, and environmental factors. It reflects the model's capacity to replicate the physiological responses of plants to their environment.

There are several process-based models that simulate tree-ring widths and/or stable isotopes (see reviews in Guiot et al., 2014; Wei et al., 2022). Such models include TREERING (Hemming et al., 2001), MuSICA (Ogee et al., 2009), 3-PG (Ulrich et al., 2019; Wei et al., 2014a), MAIDENiso (Danis et al., 2012), ISOCASTANEA (Eglin et al., 2010), ORCHIDEE (Barichivich et al., 2021), and so on. There are more models that have incorporated components for simulating stable isotopes of new photosynthate and/or leaves but not yet specifically simulate tree-ring stable isotopes; it would be convenient to upgrade these models for simulating tree-rings and expand the use of tree-ring benchmarks on improving forest modeling (see reviews of Wei et al., 2022).

One of the most widely used forest growth models, 3-PG (Physiological Processes Predicting Growth; Landsberg and Waring, 1997), can be a good representation about how tree-ring benchmarks can be used to improve forest simulations. Despite its simple structure, 3-PG simulates both tree-ring width and stable isotopes after a series of upgrades (Ulrich et al., 2019; Wei et al., 2014a; Wei et al., 2014b; Yu et al., 2024). Simulations of tree-ring stable isotopes in 3-PG have validated its descriptions of gas-exchange (Ulrich et al., 2019; Wei et al., 2014a). The 3-PG model estimates gross primary production (GPP) from absorbed photosynthetically active radiation and canopy light-use efficiency. A fixed ratio of GPP is used for respiration, and the remainder, net primary production (NPP), is distributed to leaves, stems, and roots mainly based on allometric relationships that vary by nutrition (Landsberg et al., 2003; Landsberg and Waring, 1997). The simplified Farquhar model (Farquhar et al., 1982) was used to estimate $\delta^{13}C$ in the new photosynthate based on the model's descriptions of gas exchange (i.e. GPP and canopy conductance), and the post-photosynthetic discrimination was assumed as a constant offset between the tree-ring wood and the new photosynthate as a simplification (Wei et al., 2014a). Tree-ring $\delta^{18}O$ can also be simulated by considering variations of $\delta^{18}O$ in the source water, the evaporative enrichments, and the Peclet effect in an upgraded version (Ulrich et al., 2019). Moreover, 3-PG was integrated into a recurrent neural network (RNN) designed to leverage tree-ring $\delta^{13}C$ and widths for forest model calibration (3PG-RNN; Yu et al., 2024). 3PG-RNN followed the physiological and physical mechanisms in the machine learning process; it demonstrated efficiency in calibrating key parameters associated with gas-exchange processes based on tree-ring widths and $\delta^{13}C$.

Despite the aforementioned successes in simulating tree-rings with 3-PG, there is still room for improvement. In particular, the quality of tree-ring widths and equivalent stem growth parameters (i.e. diameter growth, basal area increments (BAI), etc.) was much worse than that of simulating stable isotopes. For example, simulated tree-ring $\delta^{13}C$ were significantly correlated with observations in a study based on grand fir (*Abies grandis* Douglas ex D. Don Lindl.) stands in Northern Idaho, USA (Wei et al., 2014a), while simulated diameter growths were very different from observations. Similarly, simulated tree-ring $\delta^{13}C$ and $\delta^{18}O$ were significantly correlated with observations, while simulated BAI were not correlated with observations in a study based on Ponderosa pine in central Oregon, USA (Ulrich et al., 2019). One possible reason was that gas-exchange control over isotopic composition are relatively well understood, but carbon allocation to the stem is less well quantified.

Second, 3-PG does not have a carbon storage component, which makes it impossible to account for any carry-over effect (or memory effect, legacy effect, etc.) in tree-ring formation. The carry-over effect indicates that wood formation may use carbohydrates stored in previous years. On one hand, the tree-ring width of a year may be changed by the extra carbohydrate provided by the storage (Andreu-Hayles et al., 2022; Barbour and Song, 2014). On the other hand, ring wood formed with storage may have different stable isotopic signature than that formed using the new photosynthate (Brandes et al., 2007; Monserud and Marshall, 2001). As indicated in Wei et al., (2014a), the annual variations in simulated tree-ring $\delta^{13}C$ were larger than observations in 3-PG; carbon carryover might be expected to dampen the annual variation.

We aimed to upgrade 3-PG for better simulations for tree-ring width and $\delta^{13}C$ by enhancing its descriptions on tree physiology but keeping its simple structure and ease of parametrization. The key upgrade was adding a carbon storage component, which made 3-PG capable of accounting for the carry-over effect. We also incorporated previous modifications on 3-PG that were effective in simulating tree rings (Wei et al., 2014a; Wei et al., 2014b; Yu et al., 2024). This study could not only strengthen the effectiveness of using tree-ring information on constraining simulations of process-based models, but also improve the long-term simulations of forests.

## 2 Methods

### 2.1 Study sites

The study sites were located in the Mica Creek Experimental Watershed (MCEW) in northern Idaho, USA (47°15′N, 116°25′W) which were used in previous studies using 3-PG (Wei et al., 2014a; Wei et al., 2018; Yu et al., 2024). This region has a continental/maritime climate, characterized by wet winters and dry summers. The mean annual temperature is 4.5°C and the mean annual precipitation is 1450 mm at MCEW. Two nearly pure even-aged stands (GF1 and GF2) were selected for the study, and grand fir comprised 90% and 87% of the basal area at GF1 and GF2 respectively. Forests in MCEW mainly regenerated after logging in the 1920s and 1930s (Hubbart et al., 2007; Wei et al., 2018). Twelve variable radius plots (VRPs) were set up at each stand. Stand properties and physiological parameters were measured in 2008 and 2009, including DBH, basal area, leaf area index (LAI), tree height, stocking, litterfall, specific leaf area, and quantum yields. Sapflux data were measured at MCEW in two growing seasons (2006 and 2007) (Gag, 2009); stand-level transpiration was estimated from the sap flux, which were used for model calibrations. Detailed information can be found in Wei et al., (2014a).

### 2.2 Meteorology data

3-PG uses monthly meteorological driving data, including mean temperatures (mean daily minimum $T_{min}$; mean daily maximum $T_{max}$; and mean daily average $T_{av}$), total precipitation, solar radiation, mean daily day-time vapor pressure deficit (*VPD*), and frost days. Meteorological observations started in 1991 at MCEW. The meteorological data were extended to 1930 based on surrounding long-term meteorology stations or statistical approaches; please refer to Wei et al., (2014a) for details.

Atmospheric $CO_2$ concentration ($C_a$) and $\delta^{13}C$ of atmospheric $CO_2$ ($\delta^{13}C_a$) were used in $\delta^{13}C$ simulations, which were obtained from the Scripps $CO_2$ Program (LJO station data, https://scrippsco2.ucsd.edu; $C_a$ since 1957 and $\delta^{13}C_a$ since 2010). $\delta^{13}C_a$ of early years were estimated with an empirical function (Feng, 1998).

**2.3 Tree ring measurements**

Tree-ring samples were collected and tree-ring widths were measured in 2007 and 2008. All trees in six VRPs of each stand were sampled (42 and 30 trees at GF1 and GF2 respectively; Fig. S1). Two cores were taken in two perpendicular directions from each tree at the diameter height. Ring widths were measured, and the chronologies were cross-dated using the COFECHA program (Holmes, 1983).

Tree-ring $\delta^{13}C$ was measured for three dominant trees in each stand from 1991 to 2007 (Wei et al., 2014a). This period was chosen because reliable, local climate data were available. We removed extractives (e.g. resin, lipids, and contaminants) from the wood (Harlow et al., 2006) prior to analysis because these compounds are often much more depleted in $^{13}C$ than in structural components. All samples were analyzed for $\delta^{13}C$ at the Stable Isotope Lab of University of Idaho (Moscow, ID, USA) with an NC 2500 EA (Carlo

Erba Instruments, Milan, Italy) coupled to a mass spectrometer (Delta+IRMS, Finigan MAT, Bremen, Germany) (see Wei et al., 2014a for detail).

**2.4 3-PG Modifications**

We made two major modifications in 3-PG to better represent physiological processes, including replacing the age modifier with a tree height modifier and incorporating a carbon storage procedure. Better descriptions of tree physiology should also improve the simulations of tree rings.

**2.4.1 Tree height modifier**

We replaced the original age modifier in 3-PG with the tree height modifier and tested its effect on tree-ring simulations. The original age modifier in 3-PG describes the reduction in photosynthesis and stomatal conductance as trees age, which leads to a decrease in forest productivity and growth (Landsberg and Waring, 1997). However, it is not the age *per se* that limits forest growth. Trees gain height through time and the difficulty of transporting water to the canopy increases, which reduces stomatal conductance and photosynthesis (Hubbard et al., 1999; Koch et al., 2004; McDowell et al., 2002; Ryan and Yoder, 1997). Therefore, we incorporated the tree height of Wei et al., (2014b) to describe the size effect more physiologically. The height modifier described the stomatal responses to tree height as:

$$g_{cx} = g_{cx0} + s_g H_t \qquad (1)$$

$$k_g = k_{g0} + s_k H_t \qquad (2)$$

where $g_{cx}$ is the maximum canopy conductance (m s$^{-1}$), $g_{cx0}$ is the $g_{cx}$ of seedlings, $H_t$ is the tree height (m), $k_g$ is the constant for stomatal responses to VPD (m bar$^{-1}$; the impact of VPD is described in 3-PG as $f_{VPD}$ = exp(-$k_g$) × VPD; see Wei et al., 2014b for detail), $k_{g0}$ is the $k_g$ of seedlings, $s_g$ (m s$^{-1}$ m$^{-1}$) and $s_k$ (m bar$^{-1}$ m$^{-1}$) are the slopes of the changes in $g_{cx}$ and $k_g$ with tree height. Moreover, the original 3-PG described the water status ($\varphi$) as $\varphi$ = min($f_{VPD}, f_{SW}$)×$f_{age}$, which included the VPD ($f_{VPD}$), soil water ($f_{SW}$), and age ($f_{age}$) modifier. We modified the equation as $\varphi = f_{VPD} \times f_{SW}$ following Wei et al. (2014b); the $f_{VPD} \times f_{SW}$ is the original equation of Jarvis (1976) (please see Wei et al., 2014b for more details).

The tree height was estimated in 3-PG with an allometric equation that is used for trees in Northern Idaho (Wykoff et al., 1982) as:

$$H_t = 0.3048 * (exp(\beta_0 + \beta_1/(D/2.54 + 1)) + 4.5) \quad (3)$$

where $D$ is the mean DBH (cm), $\beta_0$ and $\beta_1$ are species-dependent constants (fitted with field measurements in MCEW; see Wei et al., 2018), 0.3048 is the conversion factor from feet to meters, and 2.54 is the conversion factor from inches to centimeters.

**2.4.2 Carbon storage pool for the stem growth**

As stored carbohydrates can be used for wood formation, especially when the photosynthesis is low (Furze et al., 2018; Huang et al., 2021; Kimak and Leuenberger, 2015), we incorporated a simple storage pool for the stem into 3-PG specifically for wood formation using stored carbohydrate (Fig.1). Note the storage for stem growth is not necessarily the storage in the stem as carbon storage in the leaf or root may also be used for wood formation. We kept the original carbon allocation algorithms in 3-PG, in which NPP is allocated to leaf, root, and stem based on growth conditions and tree size (Landsberg and Waring, 1997). However, we split the original stem biomass pool into stem mass and storage. Most carbon allocation to the stem still went to the stem mass, but a proportion of the stem allocation was diverted to the storage pool, which was not considered part of the stem mass.

Specifically, the carbon allocated to storage ($WC_{sto}$; Mg ha$^{-1}$) was assumed to be a fraction ($\eta_{sto}$; $0 \leq \eta_{sto} \leq 1$) of the carbon allocation to the stem ($\Delta W_s$, Mg ha$^{-1}$). The principles of new carbon allocated to the storage pool followed the MAIDEN model, where the amount of new carbon allocated to storage pool is a proportion of the new assimilates (Misson, 2004). The new carbon allocated to storage is defined as:

$$!!!!!!! \qquad (4)$$

As the temperature is a key factor determining carbon storage in plants (Körner, 2015; Pyl et al., 2012; Zepeda et al., 2023), we estimated $\eta_{sto}$ based on an empirical temperature-dependent function from the MAIDEN model (Misson, 2004) as:

$$!!!!!!!! \qquad (5)$$

where $k_s$ is a constant (°C) and $T_{max}$ is the daily maximum air temperature (°C). The value of $\eta_{sto}$ decreases with increasing $k_s$ and decreasing $T_{max}$ (Fig.S2).

Phenology is another key factor that determines carbon storage. It is a common phenomenon that photosynthesis begins before and continues after the cessation of radial growth in the growing season (Babst et al., 2014; Palacio et al., 2014). Photosynthesis of many coniferous trees in the northern hemisphere mostly lasts until October (Kolari et al., 2007; Kramer et al., 2000; Springer et al., 2017), however, the radial growth usually ceases before August (Rossi et al., 2013; Takahashi and Koike, 2008). We hence added a parameter to describe the time (i.e., in which month; in numerical form 1-12) of cessation in radial growth ($mthEnd$) in each year. All the allocation to stem goes to the storage instead of the stem mass after the cessation of radial growth.

We applied a simple algorithm to describe the proportion ($\eta_{use}$; $0 \leq \eta_{use} \leq 1$) of the carbon storage ($WC_{sto} \times \eta_{use}$) used in a given month. We adopted the sigmoid curve of the ISOCASTANEA model (Eglin et al., 2010) to determine $\eta_{use}$ as:

$$!!!!!!!!!!!!!! \qquad (6)$$

where $k_x$ is the maximum remobilized ratio, $k_n$ is the minimum remobilized ratio, $d_{length}$ is the mean daylength in each month (hour), and $a_{ss}$ and $d_{ss}$ determine the curvature and symmetry point of the curve respectively. This function works similarly to a switch (Fig. S3); The $\eta_{use}$ value switched from close to $k_x$ (when daylength is long) to close to $k_n$ (when day length is short) around the time when $d_{length} = d_{ss}$, and $\eta_{use} = (k_x + k_n)/2$ when $d_{length} = d_{ss}$.

The size of the stem storage pool was then simulated as a mass balance with $WC_{sto}$ as the input and $WC_{use}$ as the output as:

$$!!!!!!!!!!!!!!! \qquad (7)$$

where $WC$ and $WC_0$ are storage size in the current and previous timestep, respectively. All $WC_{use}$ is then used for the formation of stem mass in the current timestep (Fig. 1). The simple storage-related functions in this section simulated temporal dynamics of the carbon storage pool and hence made it possible to use storage for tree-ring formation and account for the carry-over effect (see results for detail).

The newly added storage pool also tracked the $\delta^{13}C$ signal. The $\delta^{13}C$ of the storage pool ($\delta^{13}C_{st}$) is important for simulating tree-ring $\delta^{13}C$ as part of stem wood is formed with stored carbon, which contains mixed $\delta^{13}C$ information and has a different $\delta^{13}C$ value than that in the new photosynthate (see next section for detail). $\delta^{13}C_{st}$ is updated at each timestep using $\delta^{13}C$ in the new photosynthate and $\delta^{13}C_{st}$ of the existing storage weighted by the $WC_{sto}$ and $WC$ the and as:

$$!!!!!!!!!!!!!!!!!!!!!! \qquad (8)$$

where $\delta^{13}C_p$ is the $\delta^{13}C$ in new photosynthate (‰), and $\delta^{13}C_{st}$ and $\delta^{13}C_{st0}$ is the stored carbon $\delta^{13}C$ (‰) in the current and previous timestep, respectively.

**2.4.3 Tree ring width simulations**

Tree-ring widths ($W_{tr}$) were estimated from simulated annual DBH increments and a bark growth adjustment factor ($k_{bark}$) (Wykoff et al., 1982) as:

$$W_{tr} = (D_n - D_{n-1})/2k_{bark} \qquad (9)$$

where $D_n$ and $D_{n-1}$ are the simulated DBH for the current and the previous time step, respectively. DBH was then calculated from stem biomass using an allometric equation (Landsberg and Waring, 1997), which was calibrated for grand fir (Wei et al., 2014a). As DBH is the diameter outside bark and tree-ring width is the radial increment inside bark, $k_{bark}$ converted out-bark diameter to in-bark diameter, and the factor 2 converted diameter to radius. $k_{bark}$ is 1.093 for grand fir based on data from northern Idaho (Wykoff et al., 1982).

We used both raw ring widths and ring-width index (RWI) to compare with model simulations. The raw tree-ring widths were directly compared with simulated tree-ring widths. However, matching only the long-term trends does not guarantee reasonable simulations for year-to-year variations in ring widths. This is because there were apparent growth trends in the raw tree-ring series: seedlings grow narrow rings, tree rings became wider as trees grew larger and peaked in the ~1960s, and the ring width decreased as the tree age after ~1960s (Fig.S1). We hence removed the age- or size-related growth trends (i.e. detrending) and used the RWI to reveal the year-to-year variations in both observed and simulated tree-ring widths. We used the Regional Curves Standardization (RCS; Esper et al., 2003) approach to detrend the observed tree-ring series. The RCS approach was chosen because all stands were even-aged and the trees shared similar age- or size-related growth patterns at each stand (i.e. very similar to "regional curves" in a region; Fig. S1). As to tree-ring simulations, because there is one ring-width series at each stand, we fitted a negative exponential curve and used it to detrend the series at each stand. The ring-width index in a given year was then estimated as simulated ring-width divided by estimated ring-width from the negative exponential curve.

We disabled the self-thinning function in 3-PG to better simulate tree-ring width following the approach of Yu et al. (2024). The self-thinning occurs when the tree number (tree ha$^{-1}$) reduces with increased tree size due to intraspecific competition; the larger trees tend to survive and small trees may die during the self-thinning process. Therefore, surviving trees for tree-ring samples should be larger trees throughout the forest growth. However, all trees are simulated as the same size, and stand biomass is evenly applied to each tree in 3-PG; the size of the "averaged tree" in 3-PG should be smaller than the sampled surviving trees for tree rings in nature. We hence force the model to simulate only surviving trees by disabling the self-thinning function and maintaining a constant stocking (408 and 326 tree ha$^{-1}$ at GF1 and GF2, respectively). This approach greatly improved the simulations of raw tree-ring widths in 3-PG. Please refer to Yu et al.(2024) for more information.

### 2.4.4 Tree ring δ¹³C simulations

Tree-ring $\delta^{13}C$ is simulated in 3-PG (Wei et al., 2014a) based on the simplified Farquhar model (Farquhar et al., 1982). The $\delta^{13}C$ in the new photosynthate ($\delta^{13}C_p$) was estimated based on the gross primary productivity (GPP; molC m$^{-2}$ s$^{-1}$) and canopy conductance for $CO_2$ ($g_c$; molC m$^{-2}$ s$^{-1}$) as:

$$\delta^{13}C_p \approx \delta^{13}C_a - a - (b-a)\frac{c_i}{c_a} \qquad (10)$$

$$c_i = c_a - \text{GPP}/g_c \qquad (11)$$

where $\delta^{13}C_a$ (‰) is $\delta^{13}C$ of the atmospheric $CO_2$, $a$ is the kinetic fractionation factor associated with the diffusion of $CO_2$ through air (4.4‰), $b$ is the net kinetic fractionation of the enzyme-catalyzed fixation of $CO_2$ by Rubisco and PEP carboxylase (27‰), $c_i$ is the internal leaf $CO_2$ mol fraction, and $c_a$ is the ambient $CO_2$ mole fraction. Note that the original 3-PG simulated canopy conductance for water vapor (m s$^{-1}$); it was first converted to the conductance for $CO_2$ (molC m$^{-2}$ s$^{-1}$) in 3-PG before applying the previous equation (Wei et al., 2014a). Tree-ring $\delta^{13}C$ ($\delta^{13}C_{tr}$) in each month is calculated as the mean of $\delta^{13}C$ from new photosynthates ($\delta^{13}C_p$) and stored carbon $\delta^{13}C$ ($\delta^{13}C_{st}$) weighted by the amount of carbon from each source (Gessler et al., 2009; Gessler et al., 2014). We applied two approaches to treat post-photosynthetic fractionation and compared them on simulating tree-ring $\delta^{13}C$ (Fig. 2). First, we added a fixed offset between the $\delta^{13}C$ of the new photosynthate/storage and the tree ring ($\varepsilon_{sp}$) to calculate $\delta^{13}C$ in the tree ring ($\delta^{13}C_{tr}$; more enriched than $\delta^{13}C_p$) to account for post-photosynthetic discriminations. The difference was measured as 1.99‰ for grand fir between tree-ring and phloem content at the study site (see Wei et al., 2014a for details). So we applied $\varepsilon_{sp}$ = 1.99‰ to the simulations. The monthly $\delta^{13}C_{tr}$ is then calculated as:

$$\delta^{13}C_{tr} = \left(\delta^{13}C_p * \frac{delWS - WC_{use}}{delWS} + \delta^{13}C_{st} * \frac{WC_{use}}{delWS}\right) + \varepsilon_{sp} \qquad (12)$$

Second, we considered two different offsets in $\delta^{13}C$ between new photosynthate ($\varepsilon_{sp1}$) or storage ($\varepsilon_{sp2}$) and tree rings. So the $\delta^{13}C_{tr}$ is calculated as:

$$\delta^{13}C_{tr} = (\delta^{13}C_p + \varepsilon_{sp1}) * \frac{delWS - WC_{use}}{delWS} + (\delta^{13}C_{st} + \varepsilon_{sp2}) * \frac{WC_{use}}{delWS} \qquad (13)$$

$\varepsilon_{sp1}$ and $\varepsilon_{sp2}$ were calibrated for the best simulations for tree-ring $\delta^{13}C$. We set $\varepsilon_{sp1}$ and $\varepsilon_{sp2}$ separately for two considerations, which is same with the set of MAIDENiso model (Danis et al., 2012). First, there could be different offsets if the wood was formed from new photosynthate or storage (Helle and Schleser, 2004; Schulze et al., 2004). In the meantime, wood tissue formed in the early growing season (earlywood) may use a higher percentage of storage than that formed in the late growing season (latewood) (Kimak and Leuenberger, 2015; Vincent-Barbaroux et al., 2019). Second, earlywood generally contains a higher percentage of lignin and less cellulose than latewood, and lignin is more isotopically depleted than cellulose (Gleixner et al., 1993; Kagawa and Battipaglia, 2022). Two phenomena synchronized with each other in a growing season; setting two offsets ($\varepsilon_{sp1}$ and $\varepsilon_{sp2}$) provided a pragmatic and effective means to include such physiological processes that were not considered in previous 3-PG versions.

Lastly, the annual $\delta^{13}C_{tr}$ was estimated from monthly $\delta^{13}C_{tr}$ weighted by the monthly increment of stem mass of each year. We then compared the observed and simulated annual tree-ring $\delta^{13}C$ to test the efficacy of the model modifications in improving tree-ring simulations.

**2.5 Model parameterization**

We parameterized the newly added functions for tree height and carbon storage. Other parameters were obtained from Wei et al., (2014a) and Ye et al. (2024), which used 3-PG in simulating the same stands.

Two parameters in the height modifier ($s_k$ and $s_g$; Eqn. 1 & 2) were calibrated to match simulated transpiration to the sapflow-based observations in 2007 (stand age 77) (Fig. S4). The height modifier should work similarly to the original age modifier in limiting the canopy conductance and transpiration after the calibration.

We calibrated the parameters for the newly added storage component to specifically improve tree-ring simulations. The approaches for the calibration are as follows. First, the best value of the phenology parameter *mthEnd* (cessation of radial growth) was determined as the month when the simulated tree-ring $\delta^{13}C$ and RWI showed the highest correlations ($R^2$) with observations. If continuous measurements of radial growth were available, such as dendrometer readings (Zweifel et al., 2021), this parameter could be directly determined at the point where radial growth ceases. Second, $k_s$ was also calibrated to reach the best tree-ring simulations. Given only a small fraction of stem allocation ($\eta_{sto}$) is typically stored and $\eta_{sto}$ is not sensitive to $k_s$ when it is small (as shown in Fig. S2), $k_s$ can be calibrated in larger steps (e.g.10 or 20). Third, the storage use function introduced four new parameters

($k_x$, $k_n$, $a_{ss}$, and $d_{ss}$), but only $k_x$ and $d_{ss}$ were important for the tree-ring simulation (see results for sensitivity test and Fig. S5); $k_n$ can be set to 0 as physiological activities, including storage use, is negligible in the cold winter of northern Idaho and the carbon use fraction was not sensitive to $a_{ss}$ (Fig. S5). Moreover, since $d_{ss}$ determines the daytime length when the carbon storage use shifts from $k_x$ to $k_n$, it can be safely calibrated by assigning it the day length of the month preceding *mthEnd* (i.e., when radial growth ends, no storage is utilized for wood formation). Consequently, we calibrated $k_x$ with the same rationale as *mthEnd* and $k_s$ to achieve the most accurate tree-ring simulations.

## 2.6 Sensitivity test

As sensitivity tests for 3-PG have been done in many previous 3-PG studies (e.g. Esprey et al., 2004; Wei et al., 2014a; Xie et al., 2020), we only tested the sensitivity of tree-ring simulations to four newly added parameters of carbon storage ($k_n$, $k_x$, $a_{ss}$, and $d_{ss}$.). We changed the parameters from their respective final values at four fixed levels (-40%, -20%, +20%, +40%) and tested how the changes would change the tree-ring simulations (Fig. S5). We used the correlations ($R^2$) between simulated and observed RWI and tree-ring $\delta^{13}C$ as the criterion of the model performance.

## 3 Results
## 3.1 Ring width series

The ring-width chronologies had strong common signals at each stand in terms of a high sensitivity, expressed population signal, and signal-to-noise ratio at each stand (see Table S2 for details). This indicated that trees grew in a relatively synchronous way within each stand, which facilitated the 3-PG modeling as the model assume all simulated trees had the same size and growth pattern.

The standard ring-width chronologies had significant 1st-order and 2nd-order autocorrelation at GF1 (i.e. the correlation between the RWI in year $n$ and that in the year $n\pm1$ and $n\pm2$) and significant 1st-order autocorrelation at GF2 (Fig. S6). The significant autocorrelation indicated that the tree growth in previous years had apparent impact on the growth of the current year.

**3.2 Model calibrations and sensitivity tests**

Two parameters for carbon allocation to storage were calibrated based on tree-ring simulations. The carbon storage parameter *mthEnd* was calibrated to 9 (i.e. September). It defines the end of the radial growth (i.e. all allocation to the stem goes to storage) and *mthEnd* = 9 provided the best simulations of tree-ring $\delta^{13}C$ and the 2nd best of RWI (*mthEnd* = 8 was marginally better) in terms of $R^2$ between simulations and observations at both stands (Fig. S7). We hence applied *mthEnd* = 9 for all simulations. Similarly, the parameter $k_s$ was determined for temperature-driven carbon allocation to storage; we calibrated it with a 10 °C increment and the calibrated values were 200 and 100 °C at GF1 and GF2 respectively. Note the large difference in $k_s$ does not translate to large differences in stem allocation; for example, ~18% and 11% of the original stem allocation were allocated to stem storage (i.e. $\eta_{sto}$ in Eqn. 5) at $T_{max}$ = 20 °C with these $k_s$ values at GF1 and GF2 respectively (see Fig. S2 for detail).

As to parameters for storage use, modeled tree-ring $\delta^{13}C$ and RWI were only sensitive to $d_{ss}$ among four newly added parameters ($k_n$, $k_x$, $a_{ss}$, and $d_{ss}$) at both stands in the sensitivity test (Fig. S5). The values for the minimum remobilized ratio $k_n$, and $a_{ss}$ (parameter that determines the curvature of Eqn. 6) were set as default ($k_n$ = 0 and $a_{ss}$ = 1.2 hour$^{-1}$; Table S1); the maximum remobilized ratio $k_x$ was calibrated to be 0.15 (unitless) and $d_{ss}$ (parameter that determines the symmetry point of Eqn. 6) was 14 (hour) at both stands based on tree-ring simulations (see Methods for detail).

We also calibrated the two offsets in $\delta^{13}C$ between new photosynthate or storage and tree rings to achieve the best simulations in tree-ring $\delta^{13}C$. The calibrated values for the offsets were 2.5‰ and 1.2‰ for $\varepsilon_{sp1}$ and $\varepsilon_{sp2}$ respectively.

**3.3 Tree-ring simulations**

Disabling the self-thinning function modified the historical trend in simulated raw ring widths. Simulated raw ring widths increased with time with self-thinning enabled, but the observed raw ring widths decreased with stand age (Fig. 3a&b). In comparison, simulations had a similar trend to the observations when the self-thinning function was disabled at both stands, and the simulations were positively correlated (Fig. 3c&d; $R^2$ = 0.87 and 0.61 at GF1 and GF2, respectively). Unless otherwise specified, all subsequent simulations disabled the self-thinning function. Besides improving simulations in raw ring widths, disabling the self-thinning process did not enhance the simulations of either RWI (Fig. 4a-d) or tree-ring $\delta^{13}C$ (Fig. 5a-d) for either of the stands.

Replacing the age modifier with the height modifier improved the simulations of tree-ring $\delta^{13}C$, but changed little in simulating ring-widths. But only a negligible increase in $R^2$ (0.01) was observed between the simulated and observed data at both stands (Fig. 3e&f). Similarly, the enhancement in RWI simulations was small, evidenced by a small increase in the $R^2$ correlation between simulated and observed RWI (by 0.03 at GF1 and 0.08 at GF2 in the 1991-2006/2008 period; Fig. 4e&f). Notable improvements were found in the simulations of tree-ring $\delta^{13}C$ (Fig. 5e&f), where $R^2$ between simulations and observations increased from 0.32 to 0.40 at GF1, and from 0.50 to 0.60 at GF2 (both $p<0.01$).

Including the stem storage pool significantly enhanced the accuracy of tree-ring simulations. Although improvements in simulating raw ring widths were only moderate in terms of $R^2$ between simulations and observations (improved by 0.03 and 0.06 at GF1 and GF2 respectively; Fig. 3g&h), there were notable improvements in simulating tree-ring $\delta^{13}C$ (Fig. 5g&h) and detrended RWI (Fig. 4g&h) with the stem storage component. The $R^2$ between simulated and observed tree-ring $\delta^{13}C$ reached 0.45 and 0.72 (p <0.01; improved by 0.05 and 0.12) with one offset in $\delta^{13}C$ between the new photosynthate and the tree ring at GF1 and GF2, respectively. Even better simulations were achieved with two offsets in $\delta^{13}C$ (between the new photosynthate/storage and the tree ring), and the $R^2$ were 0.50 and 0.73 (p <0.01; increased by 0.10 and 0.13) at GF1 and GF2 respectively (Fig. 3). Large improvements were also found in simulated RWI, where $R^2$ between simulations and observations reached 0.34 and 0.37 (p < 0.01; increased by 0.27 and 0.22) from 1991 to 2006 or from 1991 to 2008 at GF1 and GF2, respectively, which were the only significant correlations among all RWI simulations in Fig. 4.

The simulations of tree-ring widths may greatly rely on the quality of meteorological data. When the model used observed meteorological data within MCEW (after 1991) to drive the model, simulated RWI matched observations much better than those with interpolated driving data before 1991 in the simulations with carbon storage (Fig. 4g&h).

### 3.4 Monthly variation

The newly added storage component shifted the phenology of simulated stem mass increments, especially at the beginning and the end of the growing season (Fig. 6 and Fig. 7). More stem wood was formed in the spring (April and May) and less was formed in the late growing season (August-October) in the simulations with the use of stem storage than without (Fig. 6a-d). Simulated accumulated DBH increments changed slightly at the end of a year with the storage component than those without at both stands (Fig. 6e-d).

These changes in simulated stem growth (Fig. 6) were related to the newly added storage processes (Fig. 7). There were storage inputs throughout the growing season (May-October; Fig. 7a), while the storage use was only in April-August (i.e. before *mthEnd*; Fig. 7b). As a consequence, there was a net carbon storage use in April at GF1 and in April-May at GF2, and a net increase in storage from June to September at both stands (Fig. 7c&d). Moreover, the storage component shifted the pattern of monthly growth of stem mass. Increment of stem mass should have similar monthly patterns to those of the NPP or transpiration without the storage component as all allocation to stem instantly formed stem wood in the original model. In comparison, the stem wood grew with stored carbohydrate in April when NPP and transpiration were low, but ceased stem growth in September when NPP was still > 0 (Fig. 7e&f).

Simulated monthly $\delta^{13}C$ increased by 2-3‰ in the newly formed stem wood in the early growing season (April at GF1, and April-May at GF2) with the storage component at both stands; those in other months (June-August) changed little (Fig. 6g-h). Such changes in simulated monthly $\delta^{13}C$ highlighted the importance of stored carbon on monthly $\delta^{13}C$ when a large amount of storage was used in the early growing season when the NPP was low (Fig. 7b).

**3.5 Stand growth**

All approaches in this study simulated similar forest growth patterns in terms of long-term changes in DBH, stem mass, LAI, and tree height (Fig. 8), including disabling self-thinning, replacing the age modifier with the tree height modifier, and adding a carbon storage pool for the stem. However, all of these approaches simulated different growth trajectories in DBH, tree height, and LAI from the simulation in Wei et al., (2014a) with the self-thinning function enabled (Fig. 8). Simulated storage pool size generally increased with stand age, but had apparent year-to-year variations (Fig. 8e-f).

## 4. Discussion

Our modifications to 3-PG effectively improved simulations for both tree ring widths and $\delta^{13}C$. The results may represent one of the best tree-ring $\delta^{13}C$ simulations in terms of the correlations between simulations and observations (i.e. $R^2= 0.50$ at GF1 and $R^2= 0.73$ at GF2). Moreover, the model reasonably simulated both raw ring-width chronologies and year-to-year variations in detrended RWI (Fig. 3 & 4), and the $R^2$ between simulated and observed RWI were larger than 0.34 at both stands. The improvements in tree-ring simulations were achieved by strengthening the physiological basis of 3-PG, including adding the height and the carbon storage components. These modifications should make 3-PG a more reliable process-based forest growth model when tree-ring information is used for benchmarking and constraining the model.

Tree rings provide better benchmarks for simulations with the improved 3-PG than in previous model versions. In particular, as reasonably simulating $\delta^{13}C$ requires reasonably describing the gas exchange (i.e. photosynthesis, canopy conductance, and transpiration, etc.), plant stable isotopes could be a strict test for process-based forest/vegetation models by constraining and validating model simulations (Aranibar. et al., 2006; Barichivich et al., 2021; Duarte et al., 2017; Walcroft et al., 1997; Wei et al., 2014a). Among all process-based models that simulate tree-ring stable isotopes e.g. TREERING, Hemming et al., 2001; ISOCASTANEA, Eglin et al., 2010; MAIDENiso, Danis et al., 2012; etc. See review in

Wei et al., 2022), 3-PG has one of the simplest descriptions of tree physiology. Nonetheless, the tree-ring $\delta^{13}C$ component validated 3-PG's descriptions of gas exchange processes (Wei et al., 2014a; Wei et al., 2014b).

So how does the storage component improve tree-ring simulations? The calibrated parameters and simulations quantitatively revealed the impact of carbon storage on the tree-ring formation. First, the use of the stored carbohydrate damped annual variation in $\delta^{13}C$ in the simulations, which were closer to observations than simulations without storage; this was most significantly demonstrated in the years 1991-1998 at both stands (Fig. 5). Using storage for tree-ring formation in 3-PG enabled it to represent the carry-over effect (Gessler et al., 2014). Therefore, considering the carry-over effect damped simulated annual variations in $\delta^{13}C$ and hence improved the simulations. Second, using stored carbohydrates in the early spring elevated $\delta^{13}C$ values in simulated tree rings in the early growing season and reduced monthly variations in tree-ring $\delta^{13}C$. Simulated carbohydrate storage had a higher $\delta^{13}C$ than new photosynthate in the early spring and hence elevated simulated tree-ring $\delta^{13}C$ in the early spring (Fig. 6g&h). Because observed $\delta^{13}C$ of earlywood was only <1‰ more negative than that in the latewood at the study sites (Wei et al., 2014a), simulations with the storage component should be more realistic as it reduced the monthly variations from ~9‰ (without storage) to ~6‰ (with storage). Third, the newly added cessation of radial growth should make the simulations more realistic. The parameter *mthEnd* (=9 after calibration) ceased simulated radial growth in September, while the radial growth would not stop if the photosynthesis > 0 in the original 3-PG. Although we did not have phenological data to confirm the timing of *mthEnd*, the calibrated value roughly matched the general trend of the northern hemisphere. Radial growth normally ceased around August in most coniferous trees in the temperate climate of the north hemisphere (Rossi et al., 2013; Takahashi and Koike, 2008), while photosynthesis mostly continues until October (Kolari et al., 2007; Kramer et al., 2000; Springer et al., 2017). Moreover, cambium activities of trees are strongly suppressed when water potential falls below -1 MPa (Cabon et al., 2020; Muller et al., 2011), the low precipitation and low soil moisture should suppress cambium activities from July to September in the study area (Wei et al., 2014a).

The upgraded tree-ring simulations may expand the usage of 3-PG in multiple ways. First, although much improvement is needed in simulating tree-ring widths, the idea of constraining model simulations with tree-ring data can be applied more widely with tree-ring data than stable isotope data. This is because ring widths are more easily and economically obtained than tree-ring isotope data, and there are also more available ring-width data than stable isotope data in public databases. For example, the International Tree-Ring Data Bank (ITRDB) (Grissino-Mayer and Fritts, 1997) had stable isotopes measurement at only 24 sites, compared with 4200 sites with tree-ring width (Babst et al., 2017). Second, future projects may consider including the simulations of tree-ring $\delta^{18}O$ for a comprehensive assessment of 3-PG on simulating tree-rings. Although we did not simulate tree-ring $\delta^{18}O$ as we did not have observed data to compare with the model simulations at our study sites, an $\delta^{18}O$ enabled version of 3-PG is readily available and we have incorporated it in the Python version of the model we provided. Third, reasonable tree-ring simulations of the improved 3-PG made it possible to use 3-PG for climate reconstructions with the inverse modeling approach. Inverse modeling estimates unknown inputs, parameters, or model structures by using known output data or observational information (Boucher et al., 2014). Our simulations of tree rings greatly rely on the quality of meteorological data. When the model used meteorological data observed within MCEW (after 1991) to drive the model, simulated RWI matched observations much better than those with interpolated driving data before 1991 (Fig. 4). Such sensitivity on meteorological data made 3-PG an ideal tool to inversely interpolate the climatic changes based on the tree-ring simulations. To the best of our knowledge, there was only one study (Boucher et al., 2014) that applied inverse modeling to reconstruct temperature and precipitation based on a process-based forest model (MAIDENiso). Because 3-PG has a simpler structure than MAIDENiso but still provides reasonable simulations on tree rings, it should be relatively easier to adjust the climatic factors to reach better simulations in tree rings. A current project is upgrading the newly published machine learning version of 3-PG (Yu et al., 2024) for climatic reconstruction.

## 5. Conclusion

We improved the 3-PG model by adding a carbon storage pool and replacing the age modifier with a height modifier, which improved its ability to simulate both tree-ring widths and carbon stable isotopes and made it a more robust tool for forest modeling. The updated model demonstrated strong correlations between simulated and observed $\delta^{13}C$ values, with $R^2$ reaching 0.50 and 0.73 at the two study sites. Furthermore, the model's improved performance in simulating raw tree-ring widths and detrended ring-width indices underscores its enhanced physiological basis. These advancements not only strengthen the use of tree rings as benchmarks for model validation but also expand the potential applications of 3-PG in climate reconstructions and other ecological studies. The improved 3-PG model offers a more reliable framework for understanding the complex interactions between forest growth, climate, and physiological processes, providing valuable insights for forest management and climate change research.

## Acknowledgements

This project was supported by the National Natural Science Foundation of China (No., 42271048, 42430503, and 31971492). We thank Potlatch Corporation (now PotlatchDeltic) for contributing MCEW for scientific studies, where multiple field studies were initiated in the 2000s.

# Figures
**Figure 1**

A carbon storage pool for the stem was added to 3-PG for tree-ring simulations. The original carbon allocation algorithms in 3-PG remain unchanged. 3-PG uses a fixed ratio between the net primary production (NPP) and gross primary production (GPP). A proportion of NPP was allocated to the storage pool. The ratio of new carbon allocated to storage pool was determined by the monthly mean daily maximum temperature ($T_{max}$). The amount of storage use was mainly controlled by the monthly average day length.

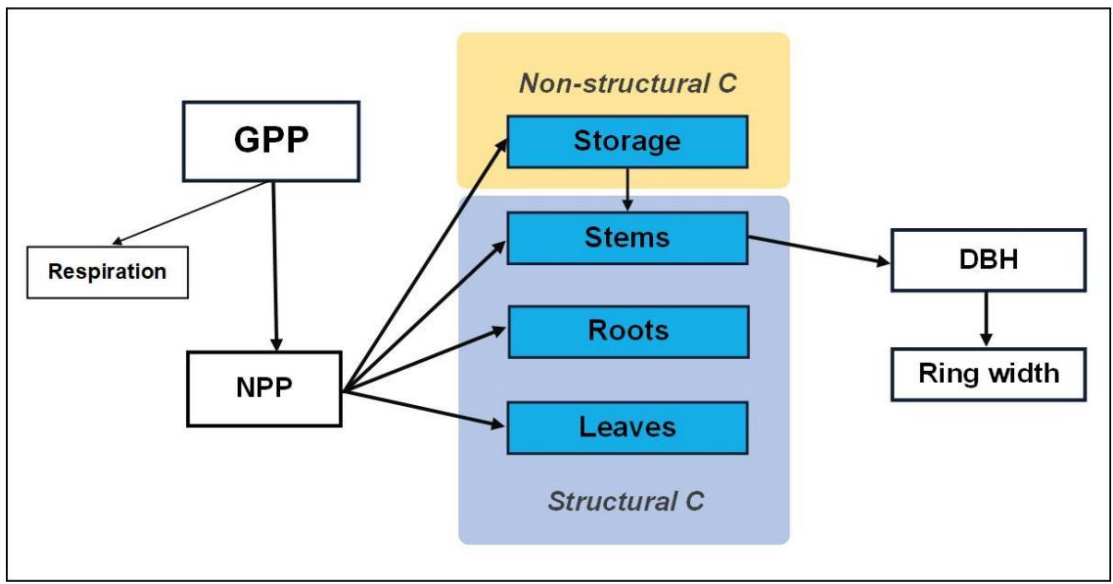

**Figure 2**

Two approaches for simulating post-photosynthetic discriminations. The first approach applied a fixed offset ($\varepsilon_{sp}$) between the tree ring and new photosynthate (same as Wei et al., 2014a) (Fig. a). Second, two separate offsets were applied between tree rings and new photosynthate ($\varepsilon_{sp1}$) or storage ($\varepsilon_{sp2}$) in the second approach (Fig. b) (see Methods for detail).

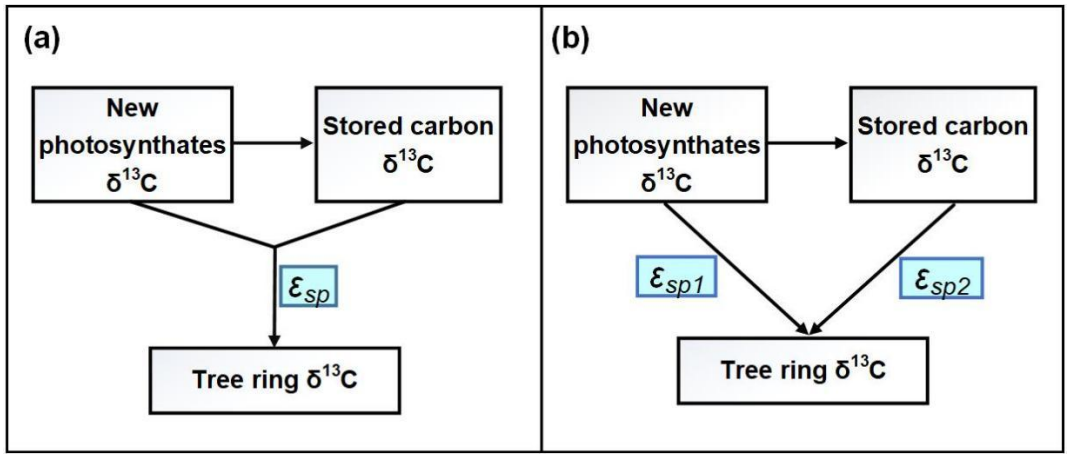

**Figure 3**

Simulated raw ring widths at stand GF1 and GF2 with different approaches, including disabling self-thinning, replacing the age modifier with tree height modifier, and adding a carbon storage pool for the stem. Observed mean ring widths of each stand are also shown. The $R^2$ and p value between observations and simulations are shown at the bottom of each figure if the simulations and observations are positively correlated.

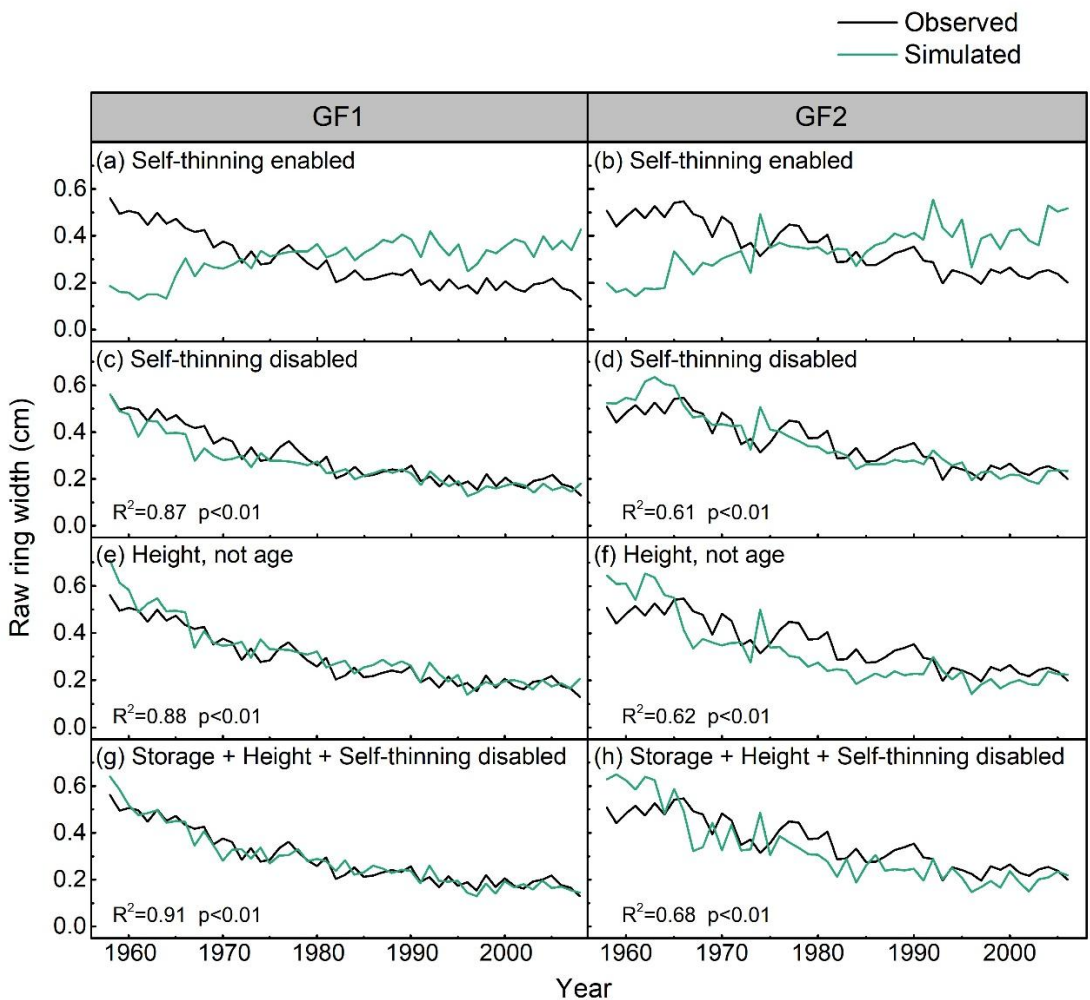

**Figure 4**

Simulated RWI at the GF1 and GF2 stand with different approaches, including disabling self-thinning (c & d), replacing the age modifier with tree height (e & f), and adding a carbon storage pool (g & h). Figures a & b show results of Wei et al., 2014 (a), which had the self-thinning function enabled. We compared the simulations with observations in three periods, 1958-1973, 1974-1990, and 1991-2008(GF1)/2006(GF2). Onsite meteorological observations started in 1991 at the study area, and tree-ring simulations should be more reasonable after 1991 than other two periods. Adding the storage component dramatically improved the RWI simulations, especially for the 1991-2006/2008 period.

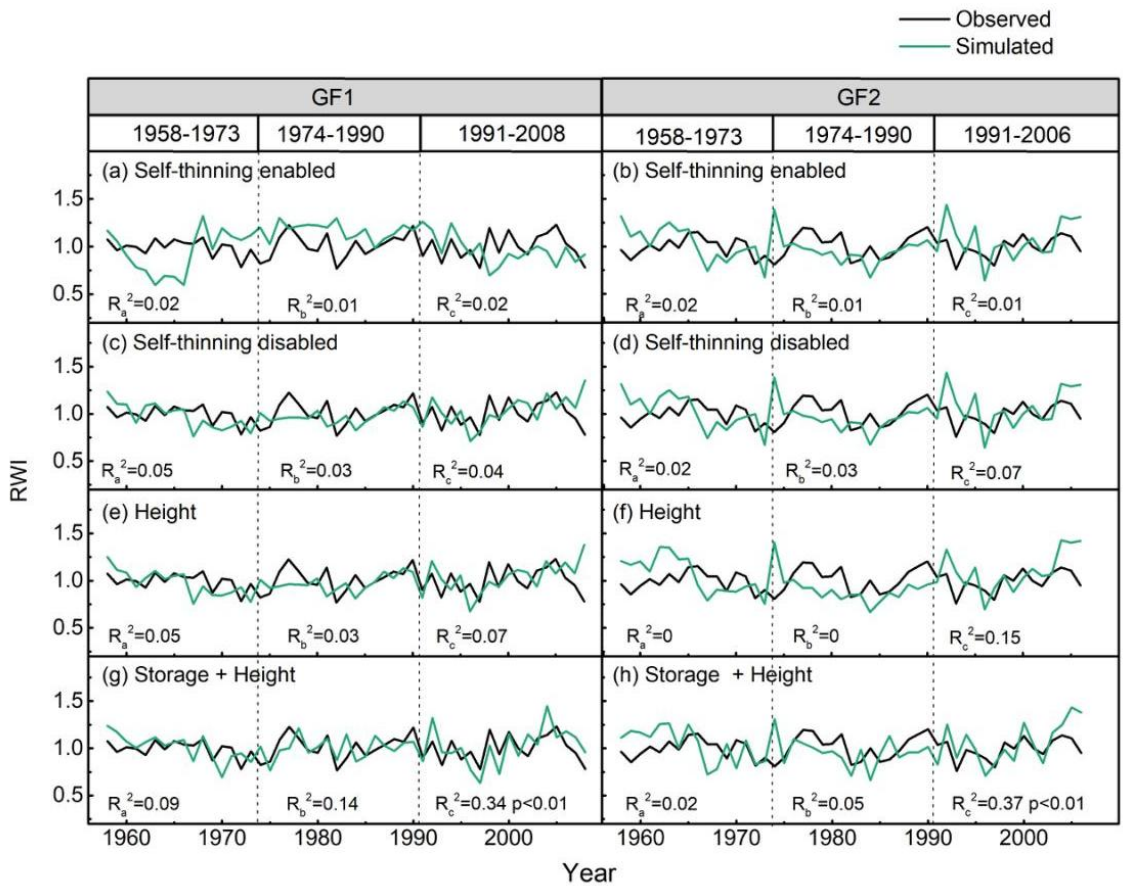

**Figure 5**

Simulated tree-ring δ¹³C from 1991 to 2007 with different approaches in the stand GF1 and GF2. Figures a & b show results of Wei et al., 2014 (a). Disabling self-thinning function did not improve the simulated tree-ring δ¹³C (c & d). In comparison, including the height effect slightly improved the δ¹³C simulations (e & f), adding carbon storage significantly improved the simulations (g & h), and simulations with two offsets (i & j) were slightly better than those with one offset in terms of the $R^2$ between simulations and observations. Error bars show SE in observations.

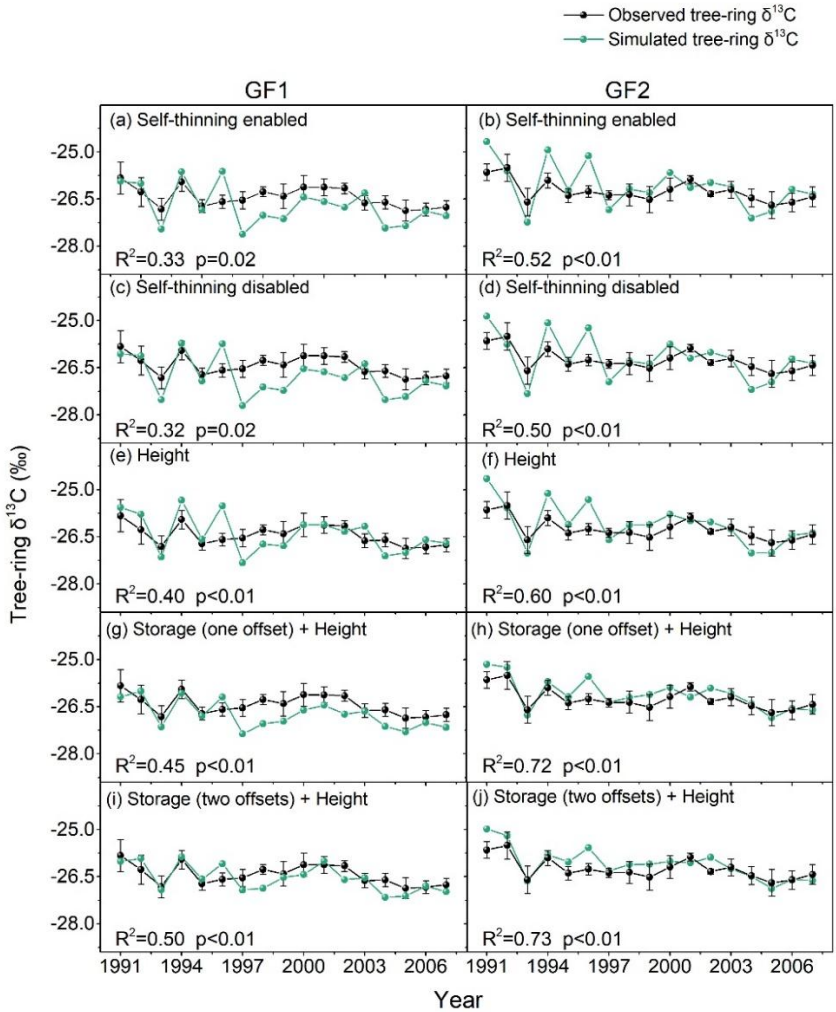

**Figure 6**

Simulated monthly variations of stem growth with and without the carbon storage component in the stands GF1 and GF2. Simulations include monthly stem mass increment (a & b), radial increment (i.e. ring width increment; c & d), accumulated DBH increment (e & f), and tree-ring $\delta^{13}C$ (g & h). The newly added storage component increased the stem growth (a-d) at the beginning of the growing season (April and May) and reduced it toward the end of the growing season (August-October). Simulated tree-ring $\delta^{13}C$ were higher around April with the storage component (g&h; only shows data in months with stem mass increment > 0). All data were averages from 1991 to 2007.

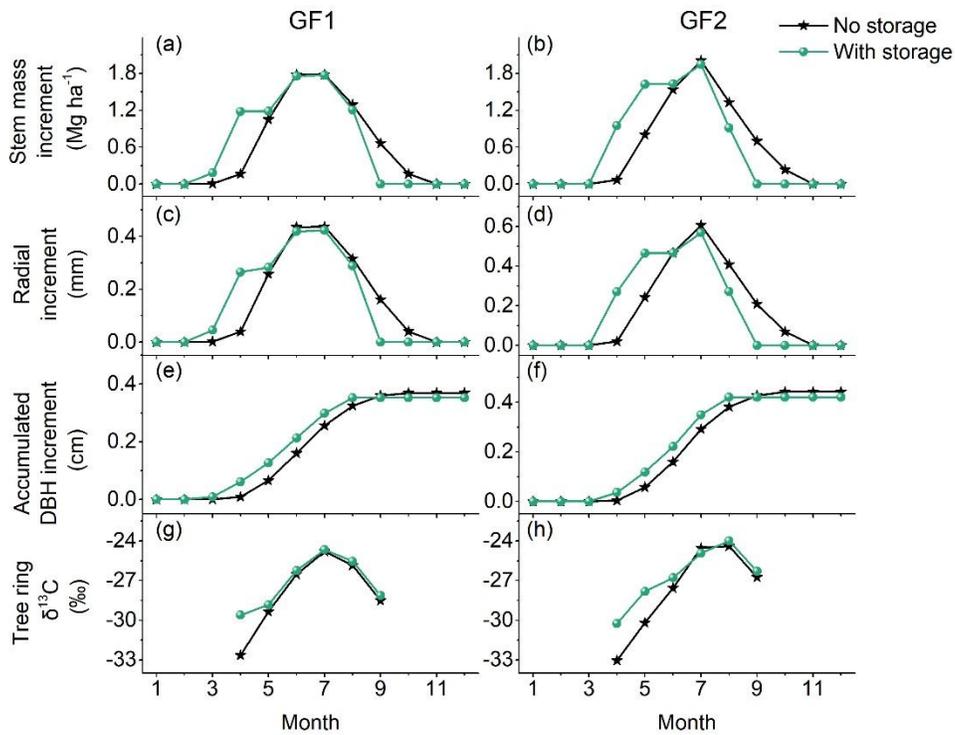

**Figure 7**

Monthly variations in simulated storage input (a)/output (b), net change (c), total storage (d), NPP and stem mass increment (e), and transpiration (f) at two stands. Observed transpirations are also shown in Fig. f, which was measured in a nearby stand with similar leaf area index in MCEW (see Fig. S4 for detail). All data were from the year 2007. Vertical dash lines in Fig. b & f show September, which is set as the end of radial growth (i.e. parameter *mthEnd*, after which all allocated carbon to stem goes to storage). Trees provided the most storage input in June, July, and September, and used the most storage for stem growth in April. With the stem storage component, the monthly stem growths no longer followed the similar pattern of NPP or transpiration.

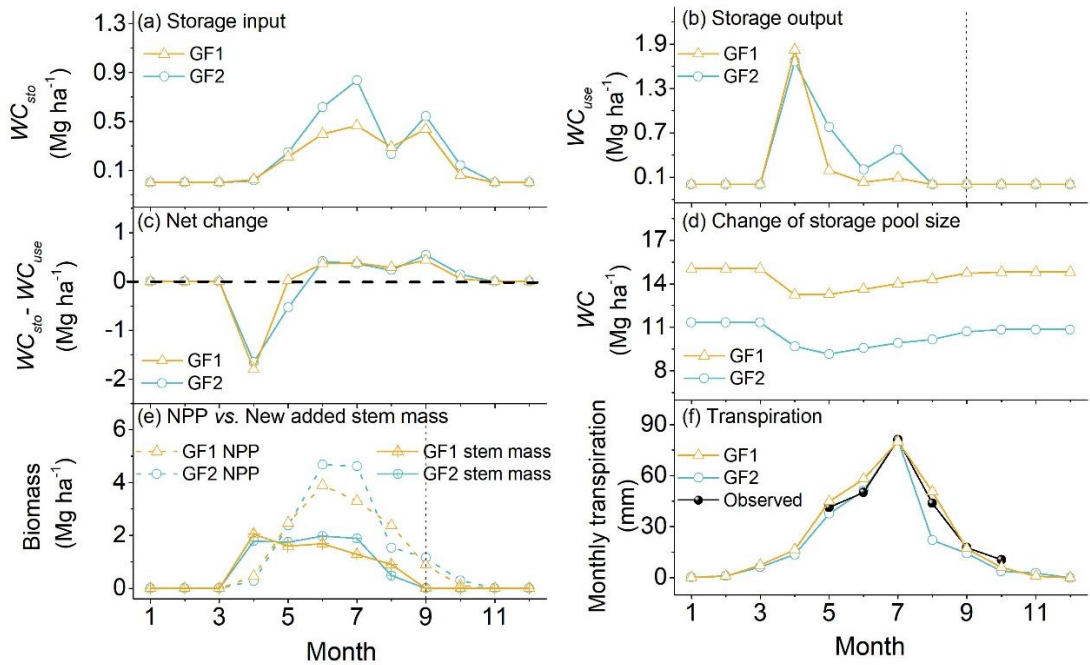

**Figure 8**

Simulated stand properties (DBH, stem dry mass, storage mass, LAI, and tree height) with different approaches, including disabling self-thinning, replacing the age modifier with the tree height modifier, and adding a carbon storage pool for the stem. All parameters, except those newly added for the storage component, were identical to Wei et al. (2014a) (i.e. "self-thinning enabled" in the figure) across different approaches (see Table S1).

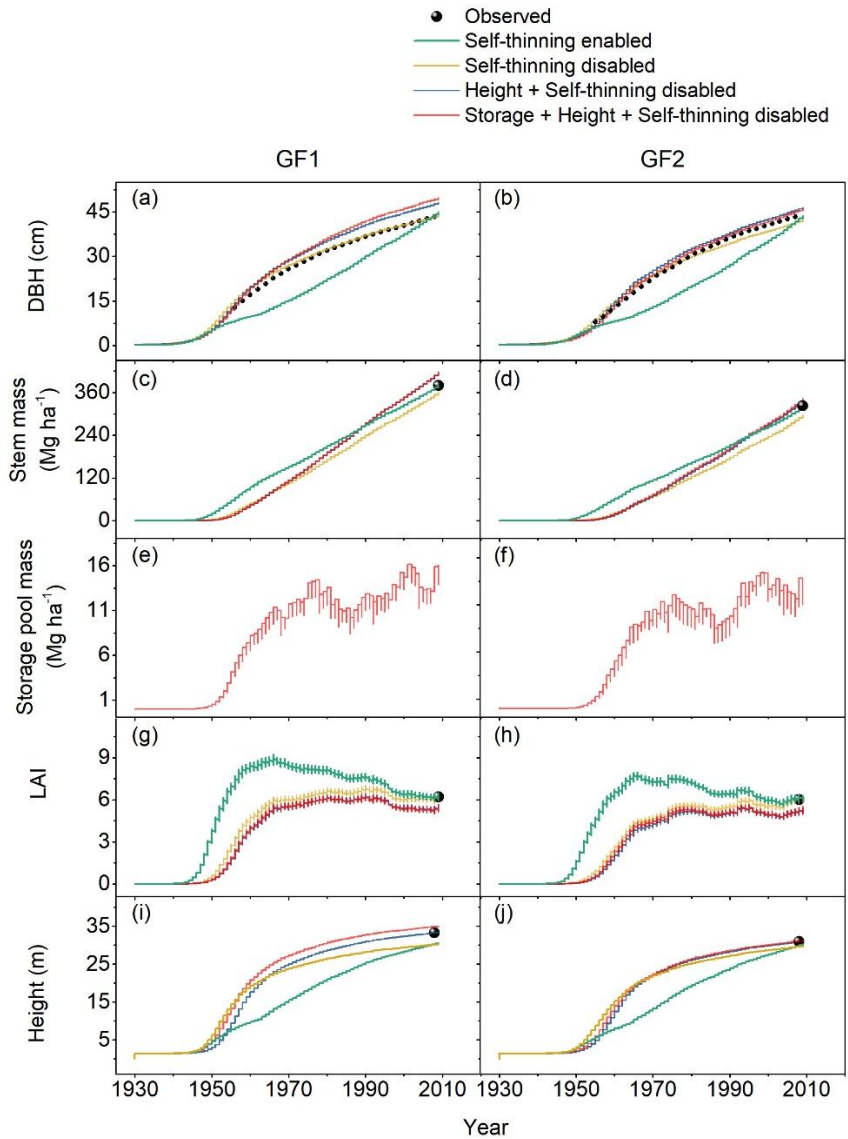

## Supplementary material

### Table S1

Parameter values and sources. Parameters are identical in two stands except those labeled as GF1 or GF2.

| Parameters | Symbol | Unit | Values | Sources |
|---|---|---|---|---|
| **Allometric relationships and partitioning** | | | | |
| Foliage: stem partitioning ratio D = 2 cm | $p_2$ | - | 0.55 | Wei et al. (2014a) |
| Foliage: stem partitioning ratio D = 20 cm | $p_{20}$ | - | 0.23 | Wei et al. (2014a) |
| Constant in the stem mass versus diameter relationship | $a_s$ | - | 0.058 | Wei et al. (2014a) |
| Power in the stem mass versus diameter relationship | $n_s$ | - | 2.549 | Wei et al. (2014a) |
| Maximum fraction of NPP to roots | $\eta_{Rx}$ | - | 0.45 | Wei et al. (2014a) |
| Minimum fraction of NPP to roots | $\eta_{Rn}$ | - | 0.25 | Default |
| Bark growth adjustment factors | $k_{bark}$ | - | 1.093 | Wykoff et al. (1982) |
| **Modifiers for photosynthesis and NPP/GPP** | | | | |
| Minimum temperature for growth | $T_{min}$ | °C | 2 | Default |
| Optimum temperature for growth | $T_{opt}$ | °C | 20 | Default |
| Maximum temperature for growth | $T_{max}$ | °C | 32 | Default |
| Days production lost per frost day | $K_F$ | day | 1 | Default |
| Ratio NPP/GPP | - | - | 0.47 | Default |
| **Soil water and fertility** | | | | |
| Moisture ratio deficit for $f_{sw}$ = 0.5 | $c_\theta$ | - | 0.7 | Default |
| Power of moisture ratio deficit | $n_\theta$ | - | 9 | Default |
| Value of 'm' when fertility rating = 0 | $m_0$ | - | 0 | Default |
| Value of '$f_N$' when fertility rating = 0 | $f_{N0}$ | - | 1 | Default |
| **Age modifier (fAge)** | | | | |
| Maximum stand age used in age modifier | $t_x$ | year | 250 | Burns et al. (1990) |
| Power of relative age in function for $f_{Age}$ | $n_{age}$ | - | 4 | Default |
| Relative age to give $f_{Age}$ = 0.5 | $r_{age}$ | - | 0.95 | Default |
| **Litterfall and root turnover** | | | | |

| Description | Symbol | Unit | Value | Source |
|---|---|---|---|---|
| Maximum litterfall rate | $\gamma_{Fx}$ | month$^{-1}$ | 0.0113 | Wei et al. (2014a) |
| Litterfall rate at t = 0 | $\gamma_{F0}$ | month$^{-1}$ | 0.001 | Default |
| Age at which litterfall rate has median value | $t_{\gamma F}$ | month | 24 | Default |
| Average monthly root turnover rate | $\gamma R$ | month$^{-1}$ | 0.015 | Default |
| **Conductance** | | | | |
| Maximum canopy conductance | $g_{cmax}$ | m s$^{-1}$ | GF1:0.0144 GF2:0.0140 | Yu et al. (2024); Calibrated |
| LAI for maximum canopy conductance | $L_x$ | - | 3.33 | Default |
| Defines stomatal response to VPD | $K_g$ | mbar$^{-1}$ | 0.08 | Calibrated |
| Canopy boundary layer conductance | $g_B$ | m s$^{-1}$ | 0.2 | Default |
| **Stem mortality** | | | | |
| Max. stem mass per tree at 1000 trees/hectare *(setting to a large number to disable mortaltiy, e.g. 9999)* | $W_{sx1000}$ | kg tree$^{-1}$ | GF1:240 GF2:190 | Wei et al. (2014a) |
| Power in self-thinning rule | $n_m$ | - | 1.5 | Default |
| Fraction mean single-tree foliage biomass lost per dead tree | $m_F$ | - | 0 | Default |
| Fraction mean single-tree root biomass lost per dead tree | $m_R$ | - | 0.2 | Default |
| Fraction mean single-tree stem biomass lost per dead tree | $m_S$ | - | 0.2 | Default |
| **Canopy structure and processes** | | | | |
| Specific leaf area at age 0 | $\sigma_0$ | m$^2$kg$^{-1}$ | 5.63 | Measured |
| Specific leaf area for mature leaves | $\sigma_1$ | m$^2$kg$^{-1}$ | 5.63 | Measured |
| Extinction coefficient for absorption of PAR by canopy | $k$ | - | 0.5 | Default |
| Age at canopy cover | $t_{cc}$ | year | 27 | Yu et al. (2024) |
| Maximum proportion of rainfall evaporated from canopy | $I_x$ | - | 0.189 | Measured |
| LAI for maximum rainfall interception | $L_{Ix}$ | - | 0 | Default |
| Canopy quantum efficiency | $\alpha_{cx}$ | molC molPAR$^{-1}$ | GF1:0.0653 GF2:0.0651 | Measured |
| **$\delta^{13}C$ submodel** | | | | |
| Temperature modifier for gc: k2 | $k2$ | - | 0.244 | Uddling et al. (2005) |
| Temperature modifier for $g_c$: $k_3$ | $k_3$ | - | 0.0368 | Uddling et al. (2005) |
| The ratio of diffusivities of CO$_2$ and water vapour in air | - | - | 0.66 | Farquhar et al. (1982) |
| $\delta^{13}C$ difference of modelled tissue and the mixture of new | $\varepsilon_{sp}$ | ‰ | 1.99 | Measured |
| photosynthate and stored carbon | | | | |
| Fractionation against $^{13}$C in diffusion through air | $a$ | ‰ | 4.4 | Farquhar et al. (1982) |
| Enzymatic fractionation by Rubisco | $b$ | ‰ | 27 | Farquhar et al. (1982) |
| **Tree height Modifier** | | | | |
| C$_0$ in the DBH-Height function | $\beta_0$ | - | 4.94586 | Calibrated |
| C$_1$ in the DBH-Height function | $\beta_1$ | - | -6.84116 | Wykoff et al. (1982) |
| Rate of $k_g$ change with tree height | $s_k$ | mBar$^{-1}$m$^{-1}$ | 0.00006 | Calibrated |
| Rate of $g_{cx}$ change with tree height | $s_g$ | m s$^{-1}$ m$^{-1}$ | -0.00001 | Calibrated |
| initial tree height for the change of $s_k$ and $s_g$ | $IniH_t$ | m | 0 | Default |
| **Newly added carbon storage pool** | | | | |
| The constant of the carbon storage function | $kS$ | °C | GF1: 200 GF2: 100 | Calibrated |
| End month of radial growth | $mthEnd$ | month | 9 | Calibrated |
| Maximum remobilized ratio | $k_x$ | - | 0.15 | Calibrated |
| Minimum remobilized ratio | $k_n$ | - | 0 | Default |
| Parameter1 relating remobilized ratio with day length | $a_{SS}$ | hour$^{-1}$ | 1.2 | Default, Eglin et al. (2010) |
| Parameter 2 relating remobilized ratio with day length | $d_{SS}$ | hour | 14 | Calibrated |
| $\delta^{13}$C difference of modelled tissue and new photosynthate | $\varepsilon_{sp1}$ | ‰ | 2.50 | Calibrated |
| $\delta^{13}$C difference of modelled tissue and stored carbon | $\varepsilon_{sp2}$ | ‰ | 1.20 | Calibrated |

**Table S2**

Descriptive statistics for two tree-ring chronologies. The mean sensitivity represents the year-to-year variations magnitude of the chronologies. The high mean series correlations ($R^2$) indicate that these tree ring series are consistent at each site and have strong common signals. A subsample signal strength (SSS) > 0.85 was used to determine the reliable period of chronologies.

| Statistical characteristics | Mean series R | Mean sensitivity | Common period | Year SSS>0.85 | EPS | PC1 (%) | SNR |
|---|---|---|---|---|---|---|---|
| GF1 | 0.639 | 0.189 | 1956-2008 | 1963 | 0.941 | 39.9 | 16.606 |
| GF2 | 0.640 | 0.163 | 1958-2006 | 1954 | 0.919 | 42.2 | 11.367 |

**Figure S1**

Raw tree-ring width series from stand GF1 and GF2. Tree-ring samples were taken from 42 trees at GF1 and 32 trees at GF2, two cores from each tree.

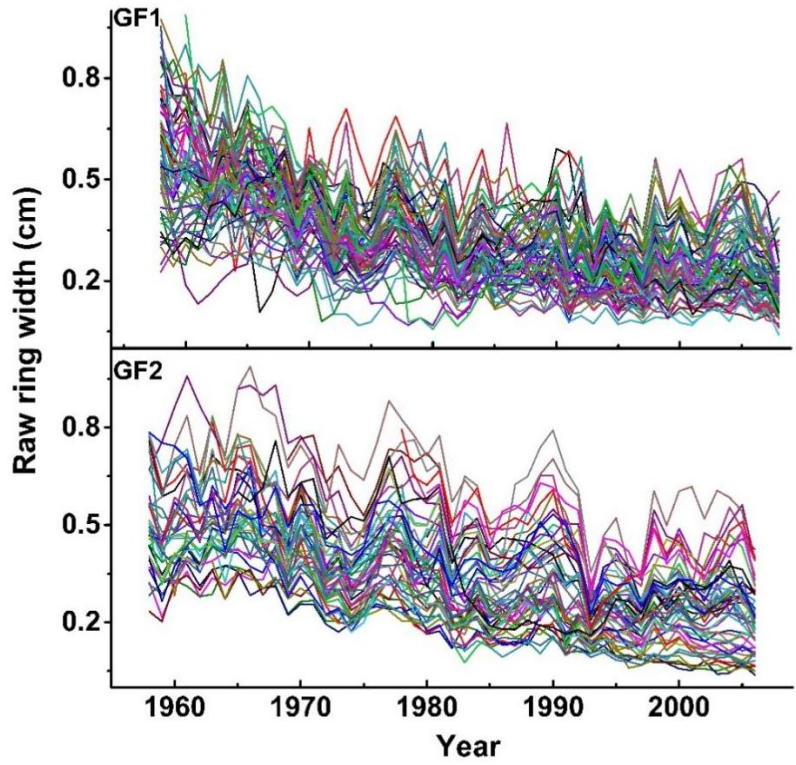

**Figure S2**

Sensitivity test of how the ratio of new assimilates allocated to the storage pool ($\eta_{c1}$; $0 \leq \eta_{c1} \leq pS$, $pS$ is the ratio of NPP allocated to the stem) varies with a parameter for carbon storage ($k_s$; unit, °C; see Methods for detail) and monthly mean daily maximum temperature ($T_{max}$).

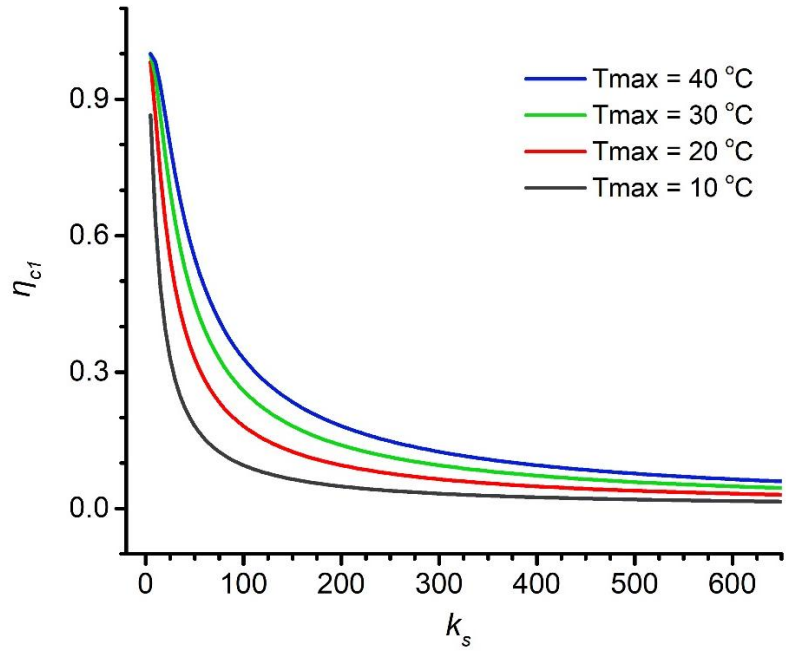

**Figure S3**

The ratio ($\eta_{c2}$; $0 \leq \eta_{c2} \leq 1$) of remobilized carbon used to stem growth varies with day length (*hour*). The equation $\eta_{c2} = \frac{k_x+k_n}{2} + \left(k_n - \frac{k_x+k_n}{2}\right) * \tanh\left(a_{ss} * (d_{length} - d_{ss})\right)$ in the ISOCASTANEA model is used to account for the ratio of stored carbon used to stem growth. The parameter $a_{ss}$ has a small effect on $\eta_{c2}$ and the parameter $d_{ss}$ has a large effect on $\eta_{c2}$. The value of $a_{ss}$ ranges from 0 to positive infinity, and the value of $d_{ss}$ ranges from 0 to 24.

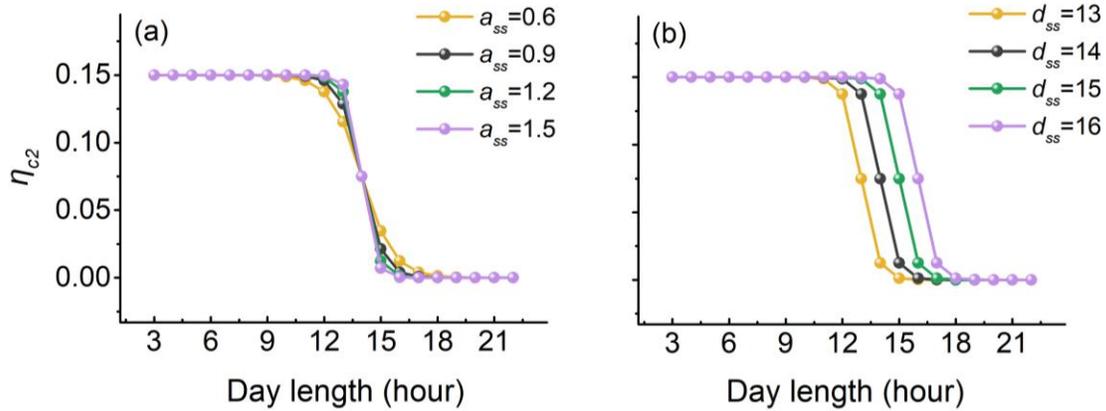

**Figure S4**

Simulated and observed monthly transpiration (a) and canopy conductance (b). Canopy conductance and transpiration which were estimated from sap flow were used for calibrating two parameters for the height effect [$s_k$ and $s_g$; $s_k$ is the rate of $k_g$ (response of canopy conductance to VPD) that changes with the tree height; $s_g$ is the rate of $g_{cx}$ (the maximum canopy conductance) that changes with the tree height]. Observations were taken from a drier stand within the watershed than GF1 and GF2, so only the data before the summer drought (May-July) were used for the calibration (see Wei et al., 2014a for detail).

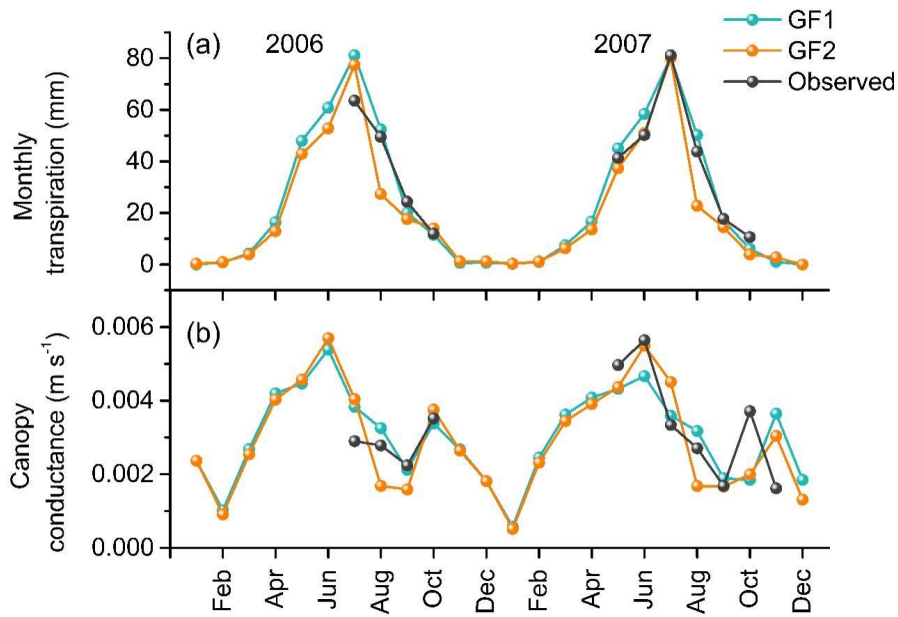

**Figure S5**

Sensitivity tests of four key parameters for carbon storage ($k_s$, $k_x$, $a_{ss}$, and $d_{ss}$) on simulating tree-ring $\delta^{13}C$ and widths. We changed parameter values by ±20% and ±40% from their respective final values, and tested how tree-ring simulations would change in terms of $R^2$ between observed and simulated tree-ring $\delta^{13}C$ and widths. We used the parameter value as the final value when it provided the highest sum of $R^2$ in both tree-ring $\delta^{13}C$ and widths simulations. The simulations were most sensitive to *dss*.

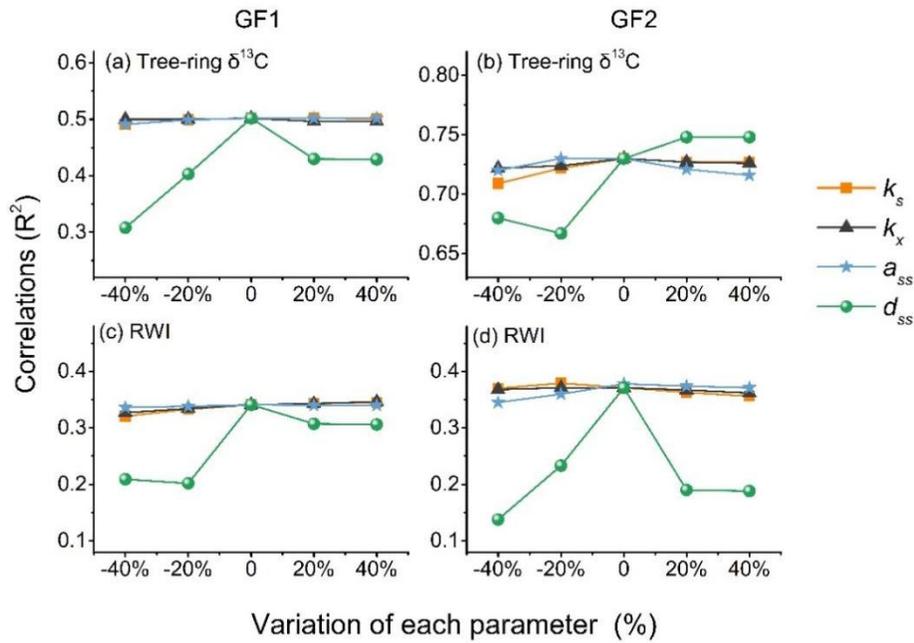

**Figure S6**

Autocorrelations in the standard chronologies at stand GF1 and GF2. The letter "n" indicated the current year, n-1 indicates the previous year of n, and n+1 indicted the next year of n, etc. The symbol * and ** indicate significant correlations at p = 0.05 and 0.01, respectively.

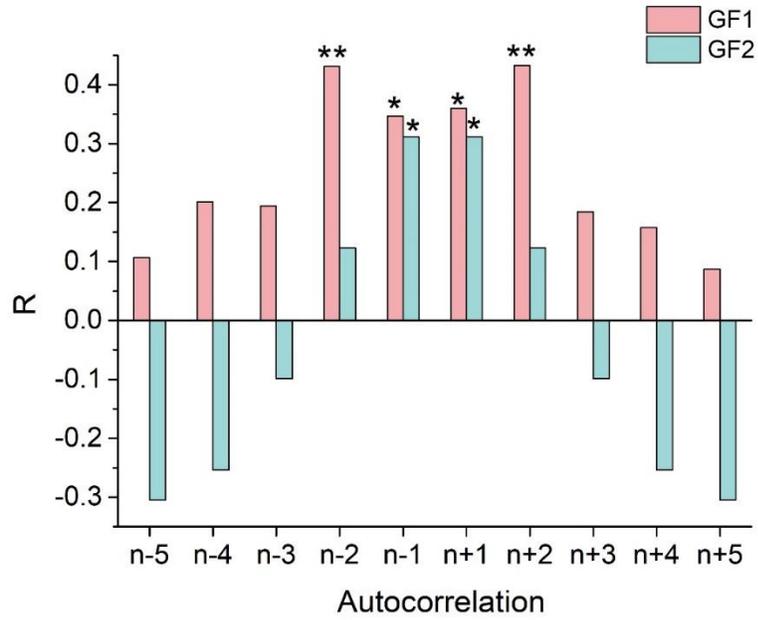

**Figure S7**

The sensitivity test of the parameter *mthEnd* on simulations of tree-ring $\delta^{13}C$ and ring-width index (RWI). The quality of the simulated tree-ring $\delta^{13}C$ and RWI was expressed as correlations ($R^2$) between simulations and observations. We set the final parameter value of *mthEnd* = 9 (vertical dash lines) as the sum of $R^2$ (RWI and $\delta^{13}C$) reached the highest among all values from July to November. Horizontal dash lines indicated $p<0.01$ if the $R^2$ is higher than this line.

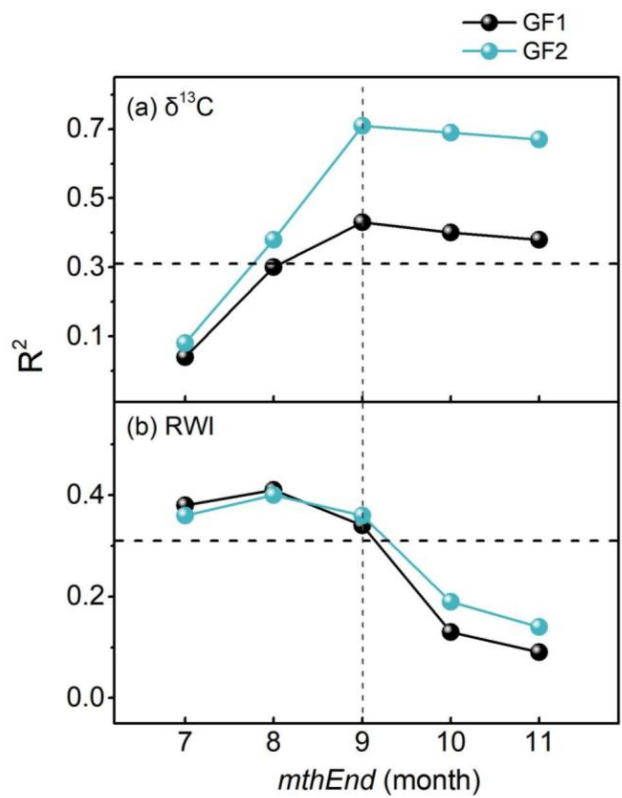